\documentclass[12pt]{article}
\usepackage{graphics}
\thicklines
\newcommand{\EQ}{\begin{equation}}
\newcommand{\EN}{\end{equation}}
\newcommand{\bea}{\begin{eqnarray}}
\newcommand{\ena}{\end{eqnarray}}
\newcommand{\eea}{\end{eqnarray}}

\def\del{\Delta}
\def\ddel{{}^\bullet\! \Delta}
\def\deld{\Delta^{\hskip -.5mm \bullet}}
\def\dddel{{}^{\bullet \bullet} \! \Delta}
\def\ddeld{{}^{\bullet}\! \Delta^{\hskip -.5mm \bullet}}
\def\deldd{\Delta^{\hskip -.5mm \bullet \bullet}}

\def\la{\langle}
\def\ra{\rangle}
\def\t{\tau}
\def\s{\sigma}
\def\p{\partial}
\def\R{\rightarrow}
\begin{document}
\begin{flushright}
\begin{minipage}{0.25\textwidth} hep-th/0008045 \\
YITP-00-44
\end{minipage}
\end{flushright}
\begin{center}
\bigskip\bigskip\bigskip
{\bf\Large{Dimensional regularization of nonlinear sigma models 
on a finite time interval}}
\vskip 1cm
\bigskip
F. Bastianelli $^a$\footnote{E-mail: bastianelli@bo.infn.it}, 
O. Corradini $^b$\footnote{E-mail: olindo@insti.physics.sunysb.edu} and 
P. van Nieuwenhuizen $^b$\footnote{E-mail: vannieu@insti.physics.sunysb.edu}  
\\[.4cm]
{\em $^a$ Dipartimento  di Fisica, Universit\`a di Bologna \\ and \\
 INFN, Sezione di Bologna\\ 
via Irnerio 46, I-40126 Bologna, Italy} \\[.4cm]
{\em $^b$ C. N. Yang Institute for Theoretical Physics \\
State University of New York at Stony Brook \\
Stony Brook, New York, 11794-3840, USA}\\
\end{center}
\baselineskip=18pt
\vskip 2.3cm

\centerline{\large{\bf Abstract}}
\vspace{.4cm}
We extend dimensional regularization to the case of compact spaces.
Contrary to previous regularization schemes employed for nonlinear
sigma models on a finite time interval (``quantum mechanical path 
integrals in curved space'') dimensional regularization requires 
only a covariant finite two-loop counterterm.
This counterterm is nonvanishing and given by
${1\over 8} \hbar^2 R$. 

\newpage


The regularization of nonlinear sigma models with higher 
dimensional target spaces but on a one dimensional worldline
(quantum mechanical path integrals in curved space) 
has a long and confusing history.
Early on, it was noticed by many authors that one obtains extra finite noncovariant
counterterms of order $\hbar^2$ in the actions for the path integral 
if one goes from the hamiltonian to the lagrangian approach.
These results were obtained in various ways: by using the Schroedinger
equation for the transition element
\cite{DeWitt:1957at},
Weyl ordering of the hamiltonian
\cite{some,Gervais:1976ws}, 
canonical point transformations in path integrals with time slicing
\cite{Gervais:1976ws}
or by making 
a change of variables at the operatorial level from field variables 
to collective coordinates and nonzero modes \cite{Tomboulis:1975gf}.
Also in standard four dimensional gauge field theories such order
$\hbar^2$ counterterms were found to be present if one chooses the 
Coulomb gauge \cite{Schwinger:1962wd,Christ:1980ku} because gauge
theories become nonlinear sigma models in this gauge.

Having fixed the counterterms in the action for the path integral
has no meaning by itself. One must also specify the regularization scheme. 
Nonlinear sigma models contain double derivative couplings so they are 
superficially divergent at the one- and two-loop levels by power counting.
In \cite{ly} it was noted that one should 
take into account the factor $\prod \sqrt{{\rm det}g_{ij}}={\rm exp}\biggl\{
{1\over 2}\delta(0){\displaystyle \int} d\t\:{\rm tr}\biggl({\rm ln}g_{ij}\biggr)
\biggr\}$ 
in the measure; exponentiating this factor by means of 
``Lee-Yang ghosts''~\cite{Bastianelli:1992be,Bastianelli:1993ct}, 
one obtains instead ${\displaystyle \int} d\t\biggl(b^ig_{ij}c^j+
a^ig_{ij}a^j\biggr)$ and the 
divergences canceled in the sums of diagrams.
Different counterterms correspond to different regularization 
schemes for these individually divergent Feynman graphs. In fact, one first 
chooses a regularization scheme 
and then determines the corresponding counterterms.

The last decade two schemes were studied in detail 
\cite{Bastianelli:1992be,Bastianelli:1993ct,deBoer:1995hv,
Schalm:1998, Bastianelli:1998jm,Bastianelli:1999jb}:
(i) mode regularization (MR) \cite{Bastianelli:1992be,Bastianelli:1993ct} 
according to which the quantum fluctuations $q(\tau)$
around a background solution $x_{cl}(\tau)$ are expanded in
a Fourier sine series cut-off at mode $N$ and all calculations 
are performed before letting $N$ tend to infinity, and
(ii) time slicing (TS) \cite{deBoer:1995hv}
according to which only $N$ variables  $q(\tau_1),\ldots,q(\tau_N)$
appear in the action at equally spaced points $\tau_i$.
In the latter case exact propagators were developed for finite $N$
and the limit $N\rightarrow\infty$ could already be implemented 
in the Feynman rules themselves.

Of course, different regularization schemes give results  
which differ by finite local counterterms. In mode regularization
these counterterms were fixed by requiring that the transition element
$\la x| {\rm exp}(-{\tau\over\hbar}H)|y\ra $
can also be obtained from a path integral with an action which differs from 
the naive action and which is fixed  by requiring that the transition
element satisfies the Schroedinger equation with the 
hamiltonian $H$.
In time slicing one also obtains a path integral representation
for $\la x| {\rm exp}(-{\tau\over\hbar}H)|y\ra $ by inserting complete sets
of position and momentum eigenstates, but here all steps are deductive and 
there is no need to impose the Schroedinger equation.
Since Feynman graphs are regulated differently it comes as no surprise
that also the counterterms are different. One finds
\EQ
\begin{array}{l}
V_{MR}={\hbar^2\over 8} R  -{\hbar^2\over 24} g^{ij} g^{kl} g_{mn}
\Gamma_{ik}^m \Gamma_{jl}^n  \\ [3mm]
V_{TS}={\hbar^2\over 8} R  +{\hbar^2\over 8} 
g^{ij} \Gamma_{ik}^l \Gamma_{jl}^k. \\
\end{array}
\EN
With these counterterms, both schemes give the same answer
corresponding to an hamiltonian $H$ proportional to the 
covariant laplacian. Thus we see that a covariant 
quantum hamiltonian in the transition amplitude requires
in both cases these noncovariant counterterms in the path integral
to obtain the same covariant answer for the transition element.
Numerous two- and three-loop calculations have confirmed  these schemes
\cite{Bastianelli:1998jm,Bastianelli:1999jb,Peeters:1999ks}.
Yet, it might simplify the calculations if a regularization scheme
were found  that only needs covariant counterterms. 
One might think of using geodesic time slicing, but the positions
of the intermediate points  $q(\tau_1),\ldots,q(\tau_N)$ would depend
on the path considered and complexities overwhelm efforts 
in this direction.

The obvious choice for regularization scheme is, of course, dimensional 
regularization, but in the past we did not succeed is using this scheme 
due to the following problems:\newline
(i) for all interesting applications one needs the action defined on a
{\bf finite} time interval. This requires a modification of the 
standard formulation of dimensional regularization such that it can be applied
to a finite time interval. This is the main problem which we solve below. Once 
this problem is solved, the calculation of the transition element and anomalies
follows relatively straightforwardly.\newline
(ii) generalizing terms such as $\dot\phi\dot\phi\dot\phi\dot\phi$ in the 
Feynman 
graphs; one must decide how to write them in $n$
dimensions (as $\partial_\mu\phi\partial_\mu\phi\partial_\nu\phi
\partial_\nu\phi$ or $\partial_\mu\phi\partial_\nu\phi\partial_\mu\phi
\partial_\nu\phi$ for example). This problem has a simple 
solution~\cite{Kleinert:1999aq} which we
use below: one starts with the action $\partial_\mu \phi\partial^\mu \phi$
in $n$ dimensions and then all Lorentz indices $\mu,\,\nu$ in all 
contractions are unambiguous.

Recently, as a test project, we considered nonlinear sigma models on a 
infinite time interval. We were inspired to return to our attempts to use
dimensional regularization for nonlinear sigma models by recent papers
by Kleinert and Chervyakov~\cite{Kleinert:1999aq} who studied a nonlinear 
sigma model with a
one-dimensional target space, and considered the same problem as Gervais and
Jevicki~\cite{Gervais:1976ws}, but using ordinary dimensional regularization.
They studied
a free particle in a box of length $d$
by replacing the confining box by a smooth convex potential
$V(x)  = {1\over 2} {m^2\over  g} \tan^2 (\sqrt{g}x)$  
which grows to infinity near the walls ($x=\pm {d\over 2}$).
The field redefinition
$x\rightarrow \varphi  = {1\over {\sqrt{g}}} \tan (\sqrt{g}x)$
was made to obtain a nonlinear sigma model with a mass term 
${1\over 2} m^2 \varphi^2$.
Using dimensional regularization it was found that both models gave the same
results ``so that there is no need for an artificial potential 
term  of order $\hbar^2$ called for by previous 
authors''~\cite{Kleinert:1999aq}. 
Of course, this refers to possible noncovariant counterterms
since a one-dimensional model cannot test counterterms proportional to $R$.
However, the interpretation ``...artificial potential term...'' of the 
results of~[1-6,8-14] may be misleading.
In general counterterms up to order $\hbar^2$
are needed in any given regularization scheme 
as they mirror in this context the ordering ambiguities
present in the canonical approach to quantum mechanics.
It would be wrong to omit them.
In the regularization schemes 
discussed before, the order $\hbar^2$ counterterms, including the 
noncovariant ones, are present and are definitely correct.
We shall find later on with our modified dimensional regularization scheme 
on a finite time interval that no noncovariant counterterms are present, 
as in \cite{Kleinert:1999aq}.  
One should view this as a property of a particular regularization scheme, 
in which the coefficients of the possible noncovariant counterterms 
happen to vanish.


The fact that no noncovariant counterterms were needed for infinite time
intervals in the one-dimensional model of Kleinert and Chervyakov and in
the $D-$dimensional model of~\cite{Bastianelli:2000pt} suggested to us to 
study dimensional regularization applied to
general nonlinear sigma models. For an infinite time interval
we indeed recently found  that one only needs a covariant counterterm 
${1\over 8} \hbar^2 R$ \cite{Bastianelli:2000pt},
but for massless nonlinear sigma models 
one must add by hand a noncovariant mass term ${1\over 2} m^2 x^2$
in order to regulate infrared divergences, and the result depends on $m$.
For the really interesting applications (to anomalies and correlation 
functions of quantum field theories) one needs a finite time interval.
In this case there are no infrared divergences and covariance can be 
maintained. In this letter we shall extend the method of dimensional 
regularization used in~\cite{Kleinert:1999aq} to a finite time interval
and show that for nonlinear sigma models one needs only a covariant 
counterterm $V_{DR}={1\over 8} \hbar^2 R$. 

The model we consider is given by the following action
\bea
S[x^i]=\int^{0}_{-1} \! \! d\tau\ {1\over 2}\, g_{ij}(x)
\dot x^i \dot x^j  .
\label{eq:model}
\eea
Decomposing the paths $x^i(\tau)$ into a classical part $x^i_{cl}(\tau)$
satisfying suitable boundary conditions, and quantum fluctuations $q^i(\tau)$
which vanish at the boundary ($q^i(-1)= q^i(0)=0$)
and decomposing the lagrangian into a free part 
$ {1\over 2} g_{ij}(0)\dot q^i \dot q^j $ plus interactions,
the propagator becomes formally
\EQ
\begin{array}{l}
\displaystyle{
\la0|T q^i(\t) q^j(\s)|0\ra = - g^{ij}(0) \Delta(\t,\s)} 
\\ [3mm]
\displaystyle{
\del(\tau,\sigma) = \sum_{n=1}^{\infty} \biggl [
- {2\over {\pi^2 n^2}} {\rm \sin}(\pi n\tau)
{\rm \sin}(\pi n\sigma)\biggr ] =
\tau (\sigma + 1 ) \theta (\tau-\sigma)
+ \sigma (\tau + 1) \theta (\sigma-\tau) .}
\end{array}
\EN

In MR one truncates the sum to $N$ modes and sends 
$N\R \infty$ at the end of the calculations, 
while in TS one uses
${\partial\over \partial \sigma} \theta(\s-\t)
=\delta(\s-\t)$ where $\delta(\s-\t)$ acts like a Kronecker delta, implying
for example that $\int \!\! \int \delta(\s-\t) \theta(\s-\t)
\theta(\s-\t) ={1\over 4}$ (and not equal to ${1\over 3}$
as one might perhaps naively expect from replacing the integrand
with ${1\over 3} {\partial\over \partial \sigma} \theta^3(\s-\t)$.
In general products of distributions are ambiguous, 
but going back to time slicing they are well defined).
With these prescriptions one can unambiguously compute loop graphs.

To extend dimensional regularization to a compact time interval
$-1 \leq \t \leq 0$ we introduce $D$ extra infinite dimensions
${\bf t}= (t^1,\ldots,t^D)$, and take
the limit $D\R 0$ at the end, as in standard dimensional regularization
\cite{'tHooft:1972fi}.
We also require translational invariance in the extra dimensions.
As action in the $D+1$ dimensions we take
\bea
S[x,a,b,c]=\int d^{D+1}t \left[{1\over 2}\, g_{ij}(x)\left(
\partial_\mu x^i \partial^\mu x^j 
+a^i a^j +b^i c^j\right)+ V_{DR}(x)\right]
\label{eq:action}
\eea
where $V_{DR}$ is the counterterm in dimensional regularization,
$t^\mu=(\t, {\bf t})$ with $\mu=0,1,\ldots,D$
and $d^{D+1}t = d\t d^{D}{\bf t }$.
The propagators for this action read
\bea
\del(t,s)=\int {d^D{\bf k}\over (2\pi)^D} \sum_{n=1}^\infty 
{-2\over (\pi n)^2+{\bf k}^2}{\rm sin}(\pi n\tau) {\rm sin}(\pi n\sigma)
{\rm e}^{i{\bf k}\cdot ({\bf t}-{\bf s})} .
\label{eq:pcr}
\eea
The coordinates ${\bf t}$ and ${\bf s}$ for the extra $D$ dimensions run 
from $-\infty$ to $\infty$, and also the $D$ continuous 
momenta $\bf k$ run from $-\infty$ to $\infty$.
This propagator satisfies the Green equation\footnote{
An alternative ansatz is suggested by writing the sines in 
(\ref{eq:pcr}) as 
$(\exp i \pi n (\t-\s) -\exp i \pi n (\t+\s) )$
and modifying it into
$(\exp i k_\mu (t^\mu-s^\mu) -\exp i k_\mu (t^\mu+s^\mu) )$, 
where $k_\mu=(\pi n, {\bf k})$.
This last expression has been used for finite temperature physics 
(J. Zinn-Justin, private communication, unpublished),
but is not suitable for our purposes as it does not satisfy the Green
equation.}
\bea
(\partial_\tau^2+\partial_{\bf t}^2)\del(t,s) =
\delta^{D+1}(s,t) =
\delta(\tau,\sigma)\delta^D({\bf t}-{\bf s})
\label{eq:eom}
\eea
where $\delta(\tau,\sigma) = 
\sum_{n=1}^{\infty} 2\, {\rm \sin}(\pi n\tau){\rm \sin}(\pi n\sigma)$
is the Dirac delta on the space of functions which vanish at 
$\t ,\s=-1,0 $.

In addition to the point particle coordinates $x^i(t)$ there are 
ghosts:  one real commuting ghost $a^i(t)$  
and two real anticommuting ghosts $b^i(t)$ and $c^i(t)$ 
\cite{Bastianelli:1992be,Bastianelli:1993ct}.
They appear in the action in the combination
 $\p_\mu x^i(t) \p^\mu x^j(t) +a^i(t) a^j(t)+b^i(t) c^j(t)$, and have  
propagators 
\EQ
\begin{array}{l}
\displaystyle{ 
\la0|T a^i(t) a^j(s)|0\ra =  g^{ij}(0) \Delta_{gh}(t,s), \ \ \ \  
\la0|T b^i(t) c^j(s)|0\ra =  -2 g^{ij}(0) \Delta_{gh}(t,s) }
\\ [3mm]
\displaystyle{ 
\del_{gh}(t,s)=\int {d^D{\bf k}\over (2\pi)^D} \sum_{n=1}^\infty 
2\, {\rm sin}(\pi n\tau) {\rm sin}(\pi n\sigma)
{\rm e}^{i{\bf k}\cdot ({\bf t}-{\bf s})} =
\delta^{D+1}(t,s) =
\delta (\t, \s) \delta^D ({\bf t} -{\bf s}) .}
\label{eq:pcrgh}
\end{array}
\EN
These ghosts arise after one integrates over the momenta in the 
path integral, 
and contribute to higher loops in exactly the same way as ghosts in gauge
theories. Although their propagators are formally equal to 
delta functions which vanish in standard dimensional regularization,
they do contribute in our case because there are no 
infrared divergences, so that the usual cancellation between infrared and 
ultraviolet divergences in $\int d^{D+1} k  =0$
does not take place.

We can now calculate loop graphs treating  the $D$ dimensional momenta 
as in ordinary dimensional regularization, and performing the sums over $n$
as in finite temperature physics.
We compute all two-loops graphs which contribute to the vacuum 
energy. For this case we have  $x^i_{cl}(\t)=0$.
We shall give details of the calculations in an example below, but first 
summarize our result in the Table~\ref{tab:2lo}, where 
we give the results for each of the diagrams which contribute to the 
two-loop vacuum energy. 
In the last column we quote the tensor 
structure of the graphs  with the shorthand notation
$\partial^2 g\equiv g^{ij}g^{kl}\partial_k\partial_l g_{ij}$, 
$\partial^j g_j\equiv g^{ik}g^{jl}\partial_k\partial_l g_{ij}$,
$\partial_k g\equiv g^{ij}\partial_k g_{ij}$ and 
$g_k\equiv g^{ij}\partial_i g_{jk}$.
We record the results for time slicing, 
mode regularization
and our version of dimensional 
regularization, respectively\footnote{
To check the statement in the caption of Table~\ref{tab:2lo}  one may use that 
$ R=\partial^2 g -\partial^j g_j -{3\over 4}\left(\partial_k g_{ij}\right)^2
+{1\over 2}\left(\partial_i g_{jk}\right)\partial_j g_{ik}
+{1\over 4}\left(\partial_j g\right)^2 
-\left(\partial_j g\right)g^j +g_j^2$
and the $\Gamma\Gamma$ terms
for TS are given by $ -{1\over 8}\Gamma\Gamma ={1\over 32}
\left (\partial_i g_{jk}\right )^2 -{1\over 16} \left (
\partial_i g_{jk}\right )\left (
\partial_j g_{ik}\right )$
while for MR one has 
$ {1\over 24}\Gamma\Gamma =
{1\over 32}
\left (\partial_i g_{jk}\right )^2 -{1\over 48} \left (
\partial_i g_{jk}\right )\left (
\partial_j g_{ik}\right )$.}. 

It is clear that there are only differences for $B_3$ and $B_4$.
The computations in DR are done by using partial 
integration to bring all integrals
in a form that can unambiguously be computed at $D\R0$. 
The various manipulation are justified in dimensional regularization.
In particular, partial integration 
is always allowed in the extra $D$ dimension 
because of momentum conservation
while it can be done in the 
finite time interval whenever there 
is an explicit function vanishing at the boundary (e.g. the propagator
of the coordinates without derivatives).
Let us use the notation 
${\partial \over \partial t^\mu }\del(t,s) ={  _\mu{\del(t,s)}}$ and 
${\partial \over \partial s^\mu }\del(t,s) =  \del_\mu(t,s) $
so that eq. (\ref{eq:eom}) yields
$  _{\mu\mu}{\del(t,s)} = \del_{gh}(t,s) =\delta^{D+1}(t,s) $.
The rule for contracting which indices with which indices follows 
from the action in (\ref{eq:action}).
We find then for  $B_4$ in dimensional regularization
\bea
&&  
B_4({\rm DR}) =
\int_{-1}^{0}  \!\!\! d\tau \! \int_{-1}^{0}  \!\!\!  d\sigma \ \ 
\ddel ~ (\deld) ~  \ddeld  
\rightarrow 
\int d^{D+1}t \int d^{D+1}s ~ ( _\mu{\del}) ~  (\del_\nu) ~ (_\mu{\del_\nu})
\cr && 
= 
\int d^{D+1}t \int d^{D+1}s ~ 
( _\mu{\del}) ~  _\mu\biggl({1\over 2}(\del_\nu)^2 \biggr )
= - {1\over 2}  \int d^{D+1}t \int d^{D+1}s ~ 
( _{\mu\mu}{\del}) ~ (\del_\nu)^2  
\cr && 
= - {1\over 2}  
\int d^{D+1}t \int d^{D+1}s ~ 
\delta^{D+1}(t,s) ~ (\del_\nu)^2  
= - {1\over 2}  \int d^{D+1}t ~ (\del_\nu)^2|_t 
\cr && 
\rightarrow  - {1\over 2} 
\int_{-1}^{0}  \!\!\! d\tau \ \ddel^2|_\tau 
= -{1\over 24} \ 
\ena 
where the symbol $ |_\t$ means that one should set $\s =\t$. Similarly
\bea
&& \hskip -.5cm B_3({\rm DR}) =
\int_{-1}^{0}  \!\!\!  d\tau \! \int_{-1}^{0}  \!\!\!  d\sigma \ \ 
\del \ (\ddeld{}^2 - \Delta_{gh}^2)
\rightarrow 
\int d^{D+1}t \int d^{D+1}s ~ \del \biggl (
( _\mu{\del_\nu}) ( _\mu{\del_\nu}) - 
( _{\mu\mu}{\del})( _{\nu\nu}{\del}) \biggr ) =
\cr &&
= \int d^{D+1}t \int d^{D+1}s ~ 
\biggl ( 
- ( _\mu{\del}) ~ (\del_\nu) ~ ( _\mu{\del_\nu}) 
- \del ~ (\del_\nu) ~ ( _{\mu\mu}{\del_\nu}) 
+ ( _\mu{\del}) ~ ( _\mu{\del}) ~ ( _{\nu\nu}{\del})  
\cr &&
+ \del ~ ( _\mu{\del}) ~ ( _{\mu\nu\nu}{\del})  \
\biggr ) 
= \int d^{D+1}t \int d^{D+1}s ~ 
\biggl ( 
- ( _\mu{\del}) ~ (\del_\nu) ~ ( _\mu{\del_\nu}) 
+ ( _\mu{\del}) ~ ( _\mu{\del}) ~ ( _{\nu\nu}{\del})  
\biggr ) 
\cr &&
= - B_4 + \int d^{D+1}t \int d^{D+1}s ~  ( _\mu{\del})^2 
\delta^{D+1}(t,s) = -B_4 +  \int d^{D+1}t ~ ( _\mu{\del})^2|_t 
= {1\over 8} .
\ena
We used the identity 
$ ( _{\mu\mu}{\del_\nu}) = (\del_{\nu\mu\mu})$ obvious from (\ref{eq:pcr}).
Moreover to compute diagrams like $A_1$ it is useful to use
an identity which can be quickly derived in $D+1$ dimensions: 
recalling that we denote with a subscript $0$ the derivative along 
the original compact time direction one has
$ (( _\mu{\del_\mu})(t,s) + ( _{\mu\mu}{\del})(t,s) )  |_{t=s} = 
{  _0{ [  ({  _0{\del(t,s)} }  )  |_{t=s}  ]} }
$.

\begin{table}[h]
\begin{center}
\begin{tabular}{|l|c|c|c|c|l|} \hline 
\emph{Integral} & 
\multicolumn{3}{c|}{\emph{Results}} &\emph{Diagram} & \emph{Tensor} \\ 
\cline{2-4}
& \emph{{\rm TS}}&\emph{{\rm MR}}&\emph{{\rm DR}} & & \emph{structure}\\ \hline

$A_1\equiv\int\del |_\tau(\ddeld+\del_{gh}) |_\tau$  & 
$-{1\over 6} $ & $-{1\over 6}$ & $-{1\over 6}$ 
&$\raisebox{-1ex}{\includegraphics*[-0.5cm,0cm][2.5cm,2.5cm]{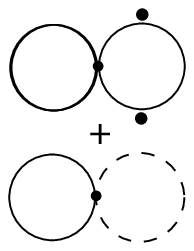}}$ & 
$-{1\over 4} \partial^2 g$  \\ \hline

$A_2\equiv\int \left (\ddel |_\tau \right )^2 $ & ${1\over 12}$ & 
${1\over 12}$ & ${1\over 12}$
&$\raisebox{-1ex}{\includegraphics*[-0.5cm,0cm][2.5cm,1.2cm]{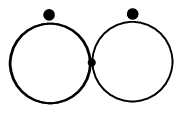}}$ 
&$-{1\over 2} \partial^j g_j$ \\ \hline

$B_3\equiv\int\int 
\del(\ddeld{}^2 -\del_{gh}^2)$ & ${1\over 4}$ & ${1\over 4}$ & ${1\over 8}$
&$\raisebox{-1ex}{\includegraphics*[-0.5cm,0cm][2.5cm,1cm]{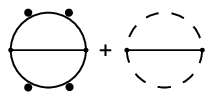}}$ &
$-{1\over 4} (\partial_i g_{jk})^2 $ \\ \hline

$B_4\equiv\int\int\left(\ddeld\right) 
\deld \left(\ddel\right) $ & 
$-{1\over 6}$
& $-{1\over 12}$ & $-{1\over 24}$ &
$\raisebox{-1ex}{\includegraphics*[-1cm,0cm][2cm,1cm]{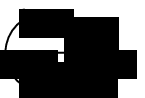}}$ &
$-{1\over 2} 
(\partial_i g_{jk})\partial_j g_{ik} $ \\ \hline

$B_5\equiv\int\int 
\ddel |_\tau\left(\ddeld\right)\deld |_\sigma$ & 
$-{1\over 12}$
& $-{1\over 12}$ & $-{1\over 12}$ 
& $\raisebox{-1ex}{\includegraphics*[-0.5cm,0cm][3cm,1cm]{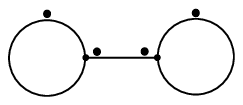}}$ & 
$-{1\over 2} g_j^2$ \\ \hline

$B_2\equiv\int\int (\ddeld
+\del_{gh}) |_\tau\deld\left(\deld |_\sigma\right)$ & ${1\over 12}$ & 
${1\over 12}$ & ${1\over 12}$
&$\raisebox{-1ex}{\includegraphics*[0cm,0cm][3cm,2.35cm]{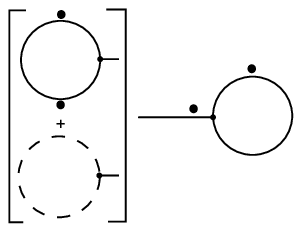}}$ & 
$ -{1\over 2} (\partial_j g)g^j$ \\ \hline

$B_1\equiv\int\int(\ddeld+\del_{gh}) |_\tau\del(\ddeld
+\del_{gh}) |_\sigma$ & $-{1\over 12}$ & $-{1\over 12}$ & $-{1\over 12}$ &
$\raisebox{-1ex}{\includegraphics*[-0.25cm,0cm][3.5cm,2.25cm]{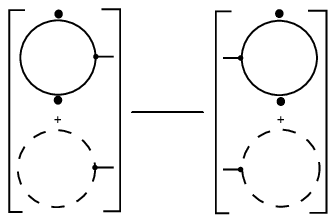}}$ 
& $ -{1\over 8} (\partial_j g)^2 $ 
\\ \hline   
\end{tabular}
\end{center}
\vspace{-0.5cm}
\caption{2-loop results with time slicing (TS), mode regularization (MR)
 and dimensional regularization (DR). 
Dots denote derivatives while hatched lined denote ghosts.
For each scheme the sum of all graphs 
and the counterterms is $- {1\over 12} R$.}
\label{tab:2lo}
\end{table}

In the calculation of $B_4({\rm DR})$ all steps are as in ordinary 
dimensional regularization, though one may question the following step 
\bea
\int d^{D+1}t \int d^{D+1}s ~  ( _{\mu\mu}{\del})~ ({\del_\nu})^2  
=  \int d^{D+1}t ~ ( {\del_\nu})^2|_t 
\label{eq:f}
\ena
where we used  that formally $ _{\mu\mu}{\del(t,s)} =\delta^{D+1}(t,s) $.
The symbol $\delta^{D+1}(t,s) $
is an analytically continued delta function, and it is not clear
that one may treat that as a regular delta function
which is defined for $D$ integer and positive.
However, we recall that the correct prescription of 
dimensional regularization is to carry out 
all integrals over spacetime at integer dimensions before
analytically continuing the momenta to $D$ dimensions.
Using this we can show by explicit calculation that 
(\ref{eq:f}) is correct.
The right hand side of eq. (\ref{eq:f})
reads 
\bea
{1\over 4}
\int d\s d^D{\bf s}  
\sum_{m_1\neq 0 } \int {d^D{\bf q}_1\over (2\pi)^D} \
\, 
{(1- e^{2 \pi i m_1 \s})\over (\pi m_1)^2 +{\bf q}_1^2}
\sum_{m_2\neq 0} \int {d^D {\bf q}_2\over (2\pi)^D}
 \,
{(1- e^{2 \pi i m_2 \s})\over (\pi m_2)^2 +{\bf q}_2^2}
(-{\bf q}_1\cdot{\bf q}_2 - \pi^2 m_1 m_2) \, I  \ 
\ena
where $I$ is unity.
The integral over $\bf s$ gives the volume of the internal space 
which can be factored out, while the integrals over
${\bf q}_1$ and ${\bf q}_2$ 
are treated with ordinary dimensional regularization which makes the sums 
over $m_1$ and $m_2$ finite for sufficiently large negative $D$.
Thus sums over modes of the finite time segment are made finite
by dimensional regularization in the internal space.
For the left hand side of equation (\ref{eq:f}),
one obtains a similar result after extracting the exponents 
containing $(\t-\s)$ from each of the three propagators, and performing
the integrals over $\bf t$ and $\t$, but now $I$ is nontrivial
\bea 
I= {1\over 2} \sum_{n\neq 0} 
(1- e^{2 \pi i n \s}) \biggl [\delta_{m_1+m_2+n,0}
+\sum_l \delta_{m_1+m_2+n-(2l+1),0} {2e^{-(2l+1) \pi i \s}\over i \pi (2l+1)}
\biggr].
\ena
We used that $\int_{-1}^0 d\t\, e^{i\pi m \t} = \delta_{m,0} +
\sum_l {2\over i\pi (2l+1)} \delta_{m-(2l+1),0}$
for any integer $m$. 
We can extend the sum over $n$ to include $n=0$.
Performing the sum over $n$
we obtain conditionally convergent series
\bea
I= {1\over 2} \biggl (1 - e^{-2 \pi i (m_1+m_2) \s} \biggr )
+ {1\over i \pi} \biggl (
S(-\s)- e^{-2\pi i (m_1+m_2) \s } S(\s)\biggr )
\ena
where $S(\s) = \sum_l {e^{(2l+1)i \pi \s} \over 2l+1}$.
The function $S(\s)$ is equal to $- {i\pi\over 2}$
for $-1\leq \s\leq 0$, and $S(-\s)= -S(\s)$, hence $I$ equals unity.
This proves (\ref{eq:f}).
 
It may be useful to compare the calculations in DR with those 
using MR and TS. Consider the integral 
$\int_{-1}^0\int_{-1}^0 \left(\ddeld\right) 
(\deld ) \left(\ddel\right) $. 
In DR  we wrote the integrand as
$( _\mu{\del}) ~  (\del_\nu) ~ (_\mu{\del_\nu}) 
=
( _\mu{\del}) ~  _\mu({1\over 2}(\del_\nu)^2 )
$,
and partially integrated the second $\mu$ derivative to obtain
\bea
&&\int d^{D+1}t \int d^{D+1}s ~ 
(- {1\over 2})  ( _{\mu\mu}{\del}) ~ (\del_\nu)^2  
=- {1\over 2}  
\int d^{D+1}t \int d^{D+1}s ~ 
\delta^{D+1}(t,s) ~ (\del_\nu)^2 = \cr &&
= - {1\over 2}  \int d^{D+1}t ~ (\del_\nu)^2|_t 
= -{1\over 24}.
\ena
In MR one can perform similar steps to arrive at 
$\int_{-1}^0\int_{-1}^0 
(- {1\over 2})   
\left(\dddel\right) (\deld )^2 $ 
but we do not set  
$\dddel=\delta(\t-\s)$
because in MR both $\delta(\t-\s)$ and $\theta(\t-\s)$
in $\deld$ 
are smeared so that one would need to work out integrals of products
of such MR regulated distributions. Instead we use the symmetry of 
$\dddel$ 
in $\t$ and $\s$ to replace $\dddel$ by $\deldd$ 
and obtain then 
\bea
B_4({\rm MR}) = -{1\over 2} \int_{-1}^0 \!\!\! d\t \! \int_{-1}^0 
\!\!\! d\s  \ 
{1\over 3}\p_\s (\deld)^3 = -{1\over 12}
\ena
where  we used that $\deld = \t +\theta(\s-\t)$.
In DR this procedure is not possible when $\mu$ is different from $\nu$. 
In the TS scheme one obtains directly without partial integration
\bea
B_4({\rm TS})= \int_{-1}^0 \!\!\! d\t \! \int_{-1}^0 \!\!\! d\s \ 
[\t + \theta(\s-\t)][\s +\theta(\t-\s)][1-\delta(\t-s)]
= -{1\over 6}
\ena
where  we used that with TS one has 
$\int_{-1}^0\int_{-1}^0 
\theta(\t-\s)\, \theta(\s-\t)\, \delta(\t-\s) ={1\over 4}$
while \linebreak
$\int_{-1}^0\int_{-1}^0 
\delta(\t-\s)\, \t \,  \theta(\t-\s) =-{1\over 4}$.

Summarizing: we have solved the problem of how to modify dimensional 
regularization such that it can be applied to finite time intervals.
Our extension of dimensional 
regularization keeps translational invariance 
in the extra $D$ dimensions. One can now regulate nonlinear sigma models 
on a finite time interval by using dimensional regularization. We were 
inspired by recent 
papers~\cite{Kleinert:1999aq} which applied standard dimensional 
regularization on an infinite time interval to a model with a mass term.
Such a mass term is also needed in more general models on an infinite time 
interval~\cite{Bastianelli:2000pt} to regularize infrared divergences, but it
breaks general covariance in target space.
Our method applies both to massless and massive models and does not break
general covariance.  
With this new method no noncovariant finite counterterms are needed. 
Other regularization schemes need such noncovariant counterterms, so 
from this perspective dimensional regularization is just another 
regularization scheme but with a simpler set of counterterms. Certainly
all regularization schemes, each with its own counterterms, yield the 
same results. In this letter we tested this for the two-loop 
vacuum energy.
Although the contributions to individual
diagrams are different,
we obtained the same results for the two-loop vacuum energy with this 
dimensional regularization and its covariant counterterms, 
$V_{DR} = {\hbar^2\over 8}R$, as previously obtained with mode regularization
and time slicing with their noncovariant counterterms.
Presently, anomalies in higher dimensions are being calculated
and we expect that our scheme will simplify such calculations.
Also applications to scattering amplitudes may benefit from this scheme. 


\end{document}